\documentclass[twocolumn]{article}
\usepackage{amsmath}
\usepackage{graphicx}
\usepackage{xcolor}
\usepackage[caption=false]{subfig}
\usepackage[numbers]{natbib}
\usepackage{siunitx}

\usepackage{ifthen}

\title{Signal anticipation and delay in excitable media: group delay of the FitzHugh-Nagumo model}
\author{Akke Mats Houben}

\begin{document}

\maketitle

\begin{abstract}

An expression for the group delay of the FitzHugh-Nagumo model in response to low amplitude input is obtained by linearisation of the cubic term of the voltage equation around its stable fixed-point. 
It is found that a negative group delay exists for low frequencies, indicating that the evolution of slowly fluctuating signals are anticipated by the voltage dynamics.

The effects of the group delay for different types of signals are shown numerically for the non-linearised FitzHugh-Nagumo model, and some observations on the signal aspects that are anticipated are stated.
\end{abstract}

\section{Introduction}
A neuronal spike is a fast transient event, lasting typically around \SI{1}{\ms}, but transmission of spikes between pairs of neurons can take between \SI{1}{} and \SI{100}{\ms} \citep{AstonJones1985, Cleland1976, Swadlow1985, Swadlow1974}, and the typical timescale of sub-threshold changes in the membrane potential of several neuron models, as fitted to experimental data, is usually set from \SI{10}{} to \SI{100}{ms} \citep{Dayan2001}. Thus we can conclude that there must exists a non-negligible delay between a neuron receiving a stimulation and the resultant action potential arriving at another neuron.

This delay seems at odds with the well-known observations that behavioural or electrophysiological responses (which both seem to involve large numbers of neurons) generally occur in the order of just a few hundred milliseconds \citep[e.g.][]{Thorpe1996, Tovee1994}. 
It is thus not surprising that rapid signal propagation between neurons and within neuronal networks has received a lot of attention \citep[e.g.][]{Delorme2003, FourcaudTromce2003, Naundorf2005, Oram1992, Panzeri1996, vanRullen2005, Tchumatchenko2011}.
It has been proposed that dynamical systems with certain properties can anticipate their input signals. 
Some (excitable) dynamical systems, such as neurons \citep[e.g.][]{Ciszak2003, Pyragiene2013} and neural systems \citep{Matias2014}, have been shown to anticipate aspects of their inputs through so-called anticipated synchronisation \citep{Voss2000}, which for excitable systems appears through canard solutions to the system \citep{KoksalErsoz2019}. 
Alternatively, model neurons with specific properties \citep{Voss2016}, as well as some electrical systems \citep[e.g.][]{Lucyszyn1993, Mitchell1997, Nakanishi2002} and certain active media \citep[e.g.][]{Bigelow2003, Brillouin1914, Chiao1993, Garrett1970, Segard1985, Sommerfeld1914, Stenner2003, Wang2000, Woodley2004}, are able to anticipate the transmission of aspects of specific signals through the existence of a negative group delay or unexpectedly high group velocities.

In this article, I show that the dynamics of a widely used model for excitable media, the FitzHugh-Nagumo model \citep{FitzHugh1961, Izhikevich2007, Nagumo1962}, near its stable fixed point $\bar{p}(\bar{v},\bar{w})$, possesses a frequency band with a negative group delay. Meaning that the dynamics of the model anticipate `sub-threshold' inputs of signals with certain characteristics within this frequency band. 
A chain of these models then seemingly respond to these types of signals before the first in the chain is good and well stimulated.

In what follows, first an illustration of negative group delay is given (section \ref{sec:ngd}), and an expression for the group delay of the FitzHugh-Nagumo model for low-amplitude input is obtained (section \ref{sec:fhn}). 
Following (section \ref{sec:res}) the effects of the group delay are demonstrated for several types of signals both numerically and conceptually. 


\section{Negative group delay}\label{sec:ngd}
Group delay $\tau(\omega)$ is defined as the additive inverse of the derivative of the phase-response $\angle \widetilde{H}(\omega)$ of a system:
\begin{equation}
    \tau(\omega) := -\frac{d}{d\omega}\angle \widetilde{H}(\omega), \nonumber
\end{equation}
and determines the time-delay of the amplitude envelope of inputs at each frequency $\omega$ \citep{Bariska2008, Mitchell1997, Woodley2004}. 

To understand the effect of a group delay, consider a signal $f(t) = g(t)e^{\mathit{i}\omega_c t}$, consisting of a sinusoid with frequency $\omega_c$, modulated by a, relative to the input, low-frequency envelope $g(t)$, being passed through a filter $\widetilde{H}(\omega) := |\widetilde{H}(\omega)|e^{\mathit{i}\phi(\omega)}$. 
For simplicity assume the filter has a flat unit amplitude response $|\widetilde{H}(\omega)| := 1$. The spectrum of $f(t)$ passed through the filter is
\begin{equation}
    \widetilde{f}(\omega)\widetilde{H}(\omega) = \widetilde{g}(\omega-\omega_c)e^{i \phi(\omega)}. \nonumber
\end{equation}
Since the modulation of the envelope is much slower than that of the sinusoidal signal, the spectrum of $G(\omega)$ is, relative to the signal, narrow around $\omega=0$. So, the output signal will contain a narrow band of frequencies, around $\omega = \omega_c$.
Approximating $\phi(\omega)$ by its Taylor series at $\omega_c$ up to first order, and substituting $\Omega = \omega-\omega_c$ gives for the inverse Fourier transform
\begin{equation}
    (f*H)(t) = e^{\mathit{i}(\omega_c t + \phi(\omega_c))} \int_{-\infty}^{\infty} \widetilde{g}(\Omega)e^{\mathit{i}\phi'(\omega_c)\Omega}e^{\mathit{i}\Omega t} d\Omega, \nonumber
\end{equation}
leading to, by the definition of group delay,
\begin{equation}
    (f*H)(t) = g(t-\tau(\omega_c))e^{\mathit{i}(\omega_c t + \phi(\omega_c))}. \nonumber
\end{equation}
Thus the envelope $g(t)$ of a signal at frequency $\omega_c$ is shifted by an amount of $\tau(\omega_c)$. Systems where $\tau(\omega)$ is negative for a range of $\omega$ are said to contain a negative group delay, and the output of these systems will anticipate the envelope of signals within this band \citep{Bariska2008, Mitchell1997, Woodley2004}.

\subsection{Group delay for filtered signals}
For a broad-band signal $f(t)$ filtered by a narrow-band filter $h_{\omega_c}(t)$ with a spectrum centered at $\omega_c$, we can show that the group-delay instead shifts the complete signal $(h_{\omega_c}*f)(t)$. Passing this signal through the same filter $H(t)$ as before gives
\begin{equation}
    ((h_{\omega_c}*f)*H)(t) = \int( \widetilde{h}_{\omega_c} \cdot \widetilde{f})(\omega)e^{\mathit{i}\phi(\omega)} e^{\mathit{i}\omega t} d\omega. \nonumber 
\end{equation}
Assuming that $\widetilde{h}_{\omega_c}(\omega)$ decays rapidly at both sides of $\omega_c$, we can use the same linear approximation as before, leading to
\begin{equation}
    e^{\mathit{i}[\omega_c t + \phi(\omega_c)]}\int(\widetilde{h}_{\omega_c}\cdot\widetilde{f})(\Omega+\omega_c)e^{\mathit{i}\phi'(\omega_c)\Omega}e^{\mathit{i}\Omega t} d\Omega, \nonumber
\end{equation}
which shows that this results in a time-shift and scaling of the filtered signal:
\begin{equation}
    ((h_{\omega_c}*f)*H)(t) = e^{\mathit{i}\phi(\omega_c)}(h_{\omega_c}*f)(t-\tau(\omega_c)). \nonumber
\end{equation}
So for signals filtered narrowly around a frequency $\omega_c$ it is the complete signal that is shifted by an amount $\tau(\omega_c)$. 

\section{... in the FitzHugh-Nagumo model}\label{sec:fhn}


The FitzHugh-Nagumo model \citep{FitzHugh1961, Nagumo1962}  describes the dynamics of the membrane potential $v$ of a neuron alongside a recovery variable $w$:
\begin{align}\label{eq:fhn}
    \frac{dv}{dt} &= v-\frac{v^3}{3} - w + I \nonumber \\
    \frac{dw}{dt} &= a(v+b-cw),
\end{align}
with three parameters: $a$, determining the timescale of the dynamics of $w$; an offset $b$; and $c$ which influences the slope of the $w$-nullcline and decay rate of $w$.

The fixed point $\bar{p}(\bar{v},\bar{w})$ of (\ref{eq:fhn}) can be easily found by solving the cubic equation $\bar{v}^3+(1/c-1)\bar{v}+b/c=I$ and plugging the resulting value for $\bar{v}$ into the equation $\bar{w}=(\bar{v}+b)/c$. 
For low-amplitude input $I$, if the fixed-point is stable we can approximate the cubic term $v^3$ by its Taylor series at $\bar{v}$, allowing the equation for the voltage evolution of (\ref{eq:fhn}) to be approximated by that of an approximated membrane potential $x=\bar{v}+\epsilon$, with governing equation
\begin{equation}\label{eq:fhnvapp}
    \frac{dx}{dt} = x - \frac{\bar{v}^3}{3} - \bar{v}^2(x-\bar{v}) - w + I + O(\epsilon^2)
\end{equation}
which is a linear first order differential equation. It is then straightforward to obtain the spectrum of the dynamics of $x$:

\begin{equation}\label{eq:fhn_spec}
    \widetilde{x}(\omega) = \frac{\widetilde{I}(\omega) + \text{DC}(\omega)}{\bar{v}^2-1+\mathit{i}\omega+\frac{a}{ac+\mathit{i}\omega}},
\end{equation}
with direct term $\text{DC} = [ab/(ac+\textit{i}\omega)+(1-1/3)\bar{v}^3]\delta(\omega)$.
If we ignore the DC term, dividing $\widetilde{x}$ by $\widetilde{I}$ leads to the transfer function $\widetilde{H}(\omega)$ of $v$ close to $\bar{v}$. 

\begin{figure}
    \subfloat[$|\widetilde{H}(\omega)|$\label{fig:mH}]{\includegraphics[width=0.5\columnwidth]{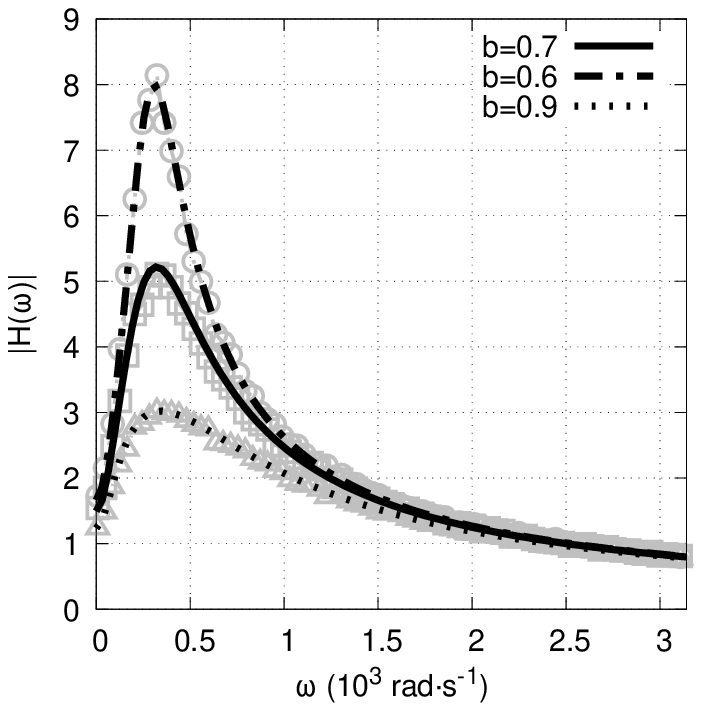}}
    \subfloat[$\angle \widetilde{H}(\omega)$\label{fig:pH}]{\includegraphics[width=0.5\columnwidth]{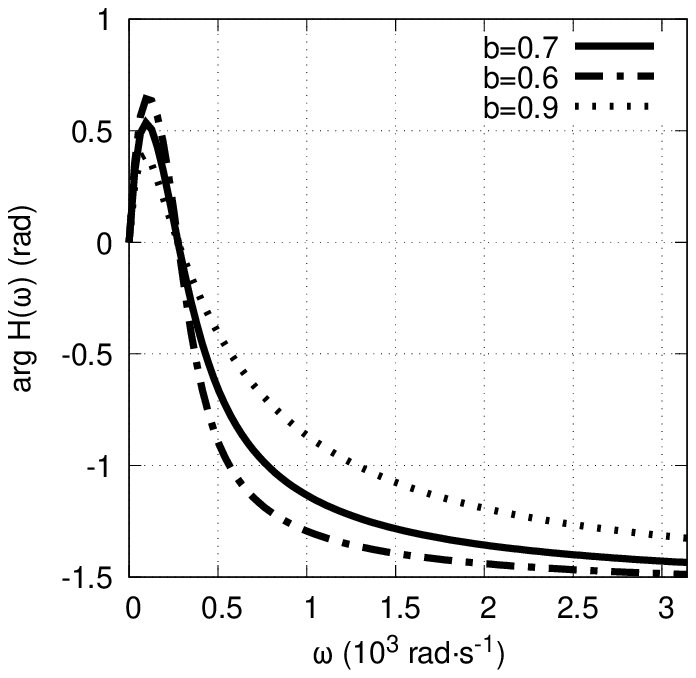}}
    \caption{\textbf{Transfer function $\widetilde{H}(\omega)$} of the linearised voltage variable of the FitzHugh-Nagumo model. a) theoretical magnitude response (black lines) and the output spectrum obtained by numerical integration of the non-linear equations (grey connected symbols). b) theoretical phase response of linearised equations. Both panels show curves for different values for $b$}\label{fig:H}
\end{figure}

Figures \ref{fig:mH} and \ref{fig:pH} show, respectively, the magnitude- and phase-transfer functions of the linearised voltage variable (solid lines) and the spectrum of the original non-linear equation obtained numerically (dashed lines) for different parameters.

The phase-spectrum is the argument of the transfer-function
\begin{equation}
    \angle \widetilde{H}(\omega) = \tan^{-1}\left(\frac{a\omega/(a^2c^2+\omega^2)-\omega}{\bar{v}^2-1+a^2c/(a^2c^2+\omega^2)}\right), \nonumber
\end{equation}
whose additive inverse differentiated with respect to $\omega$ gives the group delay

\begin{align}\label{eq:gd}
    \tau(\omega) &= -\frac{d}{d\omega}\tan^{-1}\left(\frac{ a\omega/(a^2c^2+\omega^2)-\omega}{\bar{v}^2-1+a^2c/(a^2c^2+\omega^2)}\right) \nonumber \\
    &= -\frac{\frac{a^3c-2a((\bar{v}^2-1)+ac)\omega^2}{(a^2c^2+\omega^2)^2}+\frac{a(\bar{v}^2-1)-a^2c}{a^2c^2+\omega^2}-(\bar{v}^2-1)}{[\omega a/(a^2c^2+\omega^2)-\omega]^2 + [(\bar{v}^2-1)+a^2c/(a^2c^2+\omega^2)]^2}.
\end{align}

In order for (\ref{eq:gd}) to be negative we need
\begin{equation}\label{eq:ineq}
    A\omega^4 + B\omega^2 + C < 0,
\end{equation}
where
\begin{align}
    &A := (\bar{v}^2-1) \nonumber \\
    &B := (\bar{v}^2-1)(2a^2c^2 + a) + 3a^2c  \nonumber \\ 
    &C := (\bar{v}^2-1)(a^4c^4-a^3c^2) + a^4c^3 - a^3c. \nonumber 
\end{align}

The l.h.s. of (\ref{eq:ineq}) has a real positive root only if $B^2-4AC \geq 0$ 
given by
\begin{equation}
    \omega_0 = \sqrt{\left|\frac{\sqrt{B^2-4AC}-B}{2A}\right|}. \nonumber
\end{equation}
Using the slope of (\ref{eq:ineq}) at this zero-crossing
\begin{equation}
    \left. \frac{d}{d\omega}A\omega^4+B\omega^2+C\right\rvert_{\omega = \omega_0} = 4A\omega_0^3 + 2B\omega_0, \nonumber
\end{equation}
noting that $0<\omega_0^2<a$ and that for stable fixed-points $\bar{v}^2-1+ac>0$, we obtain the result that (\ref{eq:fhnvapp}) has a frequency band with negative group delay for $\omega<\omega_0$.

The group delay is maximally negative for $\omega=0$ and increases monotonically to $\omega=\omega_0$, meaning that the envelope of signals with frequencies $0<\omega<\omega_0$ are transmitted with a negative delay.
For high frequencies (\ref{eq:gd}) tends to the limit
\begin{equation}
    \lim_{\omega\to\infty}\tau(\omega) = 0^+, \nonumber
\end{equation}
thus high frequencies are transmitted without significant delay or anticipation.
In fact, for the canonical case in which the timescales of $v$ and $w$ are sufficiently separated, $a\ll1$, the group delay approaches this limit fast for $\omega>1$. The group delay is then maximal for $\omega_0<\omega<1$.
\begin{figure}
    \includegraphics[width=.9\columnwidth]{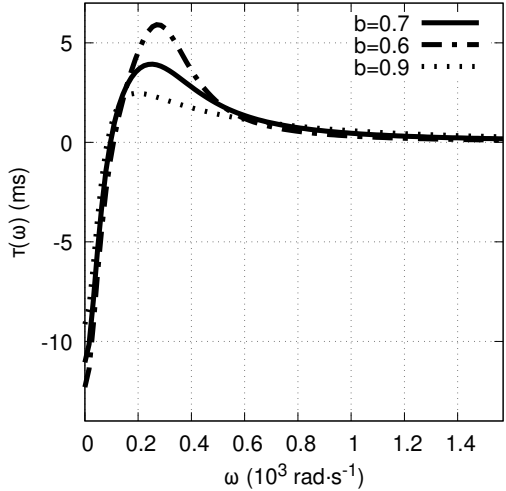}
    \caption{\textbf{Group delay function $\tau(\omega)$} of the voltage variable of the linearised FitzHugh-Nagumo model for different values for parameter $b$}\label{fig:gd}
\end{figure}
Figure \ref{fig:gd} shows the group delay function $\tau(\omega)$ for some different parameters.

\section{Anticipation of signals}\label{sec:res}
Group delay affects the envelope of band-limited signals, so the types of signals for which the group delay of the membrane potential will have an effect are the modulations of `constant' band-limited signals.
The linear approximation of $\phi(\omega)$ at $\omega=\omega_c$ during the illustration of the group delay implies, as stated before, that the frequency spectrum of the modulation needs to have a low-pass characteristic, so that the spectrum of the modulation is narrow around  $\omega=0$. 

In the following the anticipation of different types of signals by the voltage variable of the non-linearised model neuron are shown numerically. In order to differentiate between the effects of the phase delay $-\phi(\omega)/\omega$ and the group delay the simulations are carried out with a chain of $17$ model neurons, thus leading to a total expected group delay of $17\tau(\omega)$. 
In each run the first in the chain of neurons receives an input $I_0(t)$, and the subsequent neurons receive an input directly from the membrane potential of each previous neuron
\begin{equation}
    I_i(t) = \eta[v_{i-1}(t)-v_0], \text{ for $i=1,2,3,\ldots$} \nonumber 
\end{equation}
with coupling strength $\eta$. For each simulation all neurons have parameter values $a=0.08$, $b=0.7$ and $c=0.8$, leading to a negative group delay for frequencies below $ \omega_0/(2\pi) = \nu_0 \approx$\SI{15.14}{\hertz} (c.f. fig.\ref{fig:gd}), and $\eta = 0.95 |H(\omega)|^{-1}$.

Since $\tau(\omega)$ is not flat for $\omega < 1$, considerable frequency smearing is expected even for narrow-band signals in this range. Therefore, in the following numerical examples it is not expected that the magnitude of the observed time-shift corresponds absolutely to the theoretical prediction but, as will be shown, the time-shifts correspond qualitatively to the shape of $\tau(\omega)$ and in most cases agree well on the magnitude of the time-shift as well.

\subsection{Wave pulses}
The classical way to show the effects of group delay is to use wave pulses: sinusoidal signals of different frequencies modulated by a windowing function. These signals would correspond to short oscillatory bursts, which could indicate or establish transient coordination of the activity of otherwise independent elements \citep{Kuramoto1975, Strogatz2000, Winfree1967}, which is proposed to underly processes in the brain \citep[e.g.][]{Buzsaki2006, Gray1989, Varela2001}. 
In a more general sense signals of this type are amplitude modulated signals, in which the amplitude of a (high frequency) carrier signal is modulated by a lower frequency signal which is to be transmitted.

The windowing function used in the following is a Gaussian pulse
\begin{equation}
    g(t) = e^{-\alpha t^2}\nonumber
\end{equation}
of width $\sigma$, which has a spectrum of the form
\begin{equation}
    \widetilde{g}(\omega) =\sqrt{\frac{\pi}{\alpha}}e^{-\frac{\omega^2}{4\alpha}}, \nonumber
\end{equation}
which is centered at $\omega=0$ and whose magnitude $|\widetilde{g}(\omega)|$ has a fast roll off depending on $\alpha$. Thus for wide enough Gaussian pulses, leading to small values for $\alpha$, the linear approximation of the phase-spectrum should hold.

\begin{figure}
    \subfloat[$\omega_c=\omega_0/2$\label{fig:wavepulse_ngd}]{\includegraphics[width=.5\columnwidth]{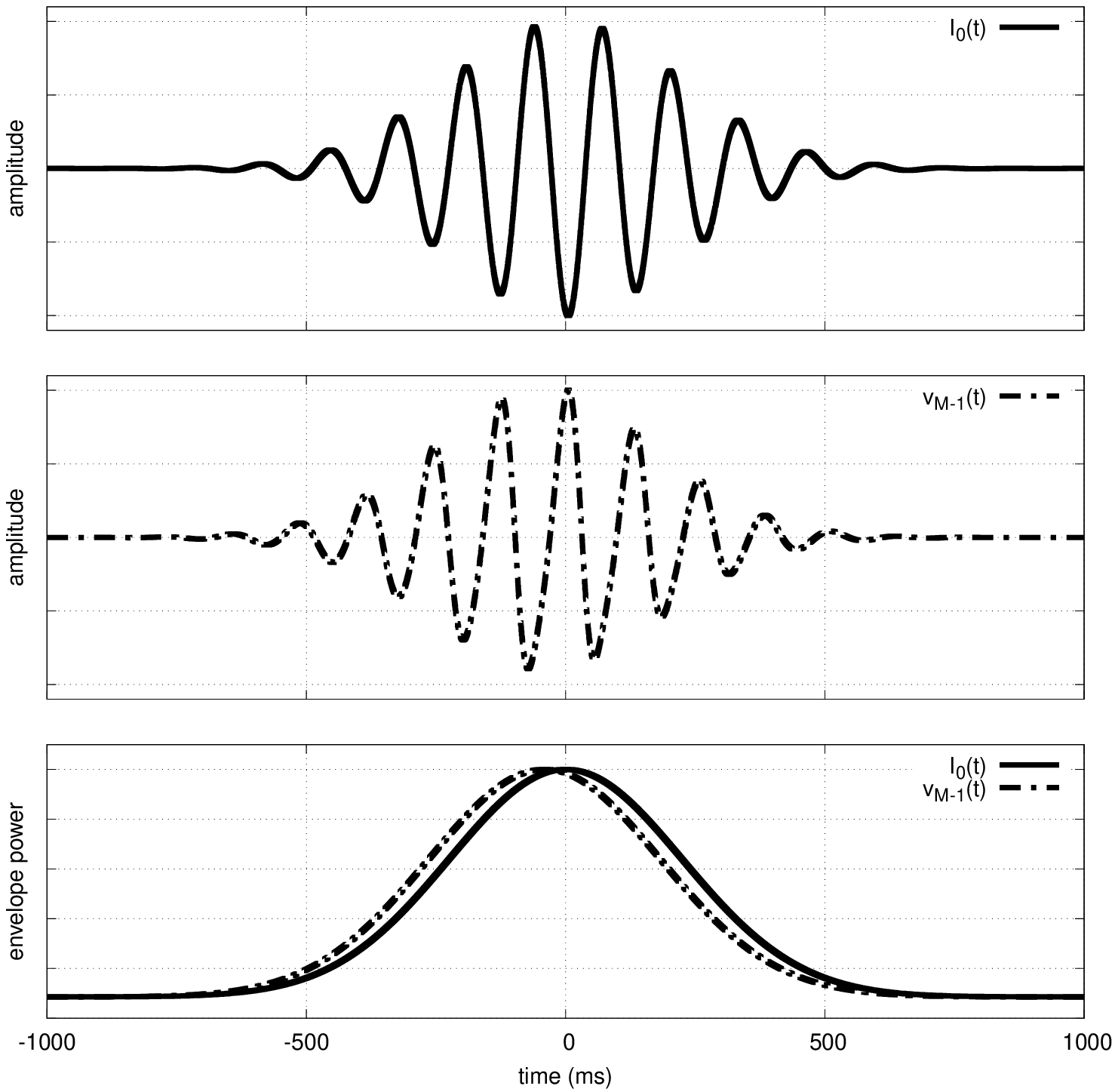}}
    \subfloat[$\omega_c=2\omega_0$\label{fig:wavepulse_pgd}]{\includegraphics[width=.5\columnwidth]{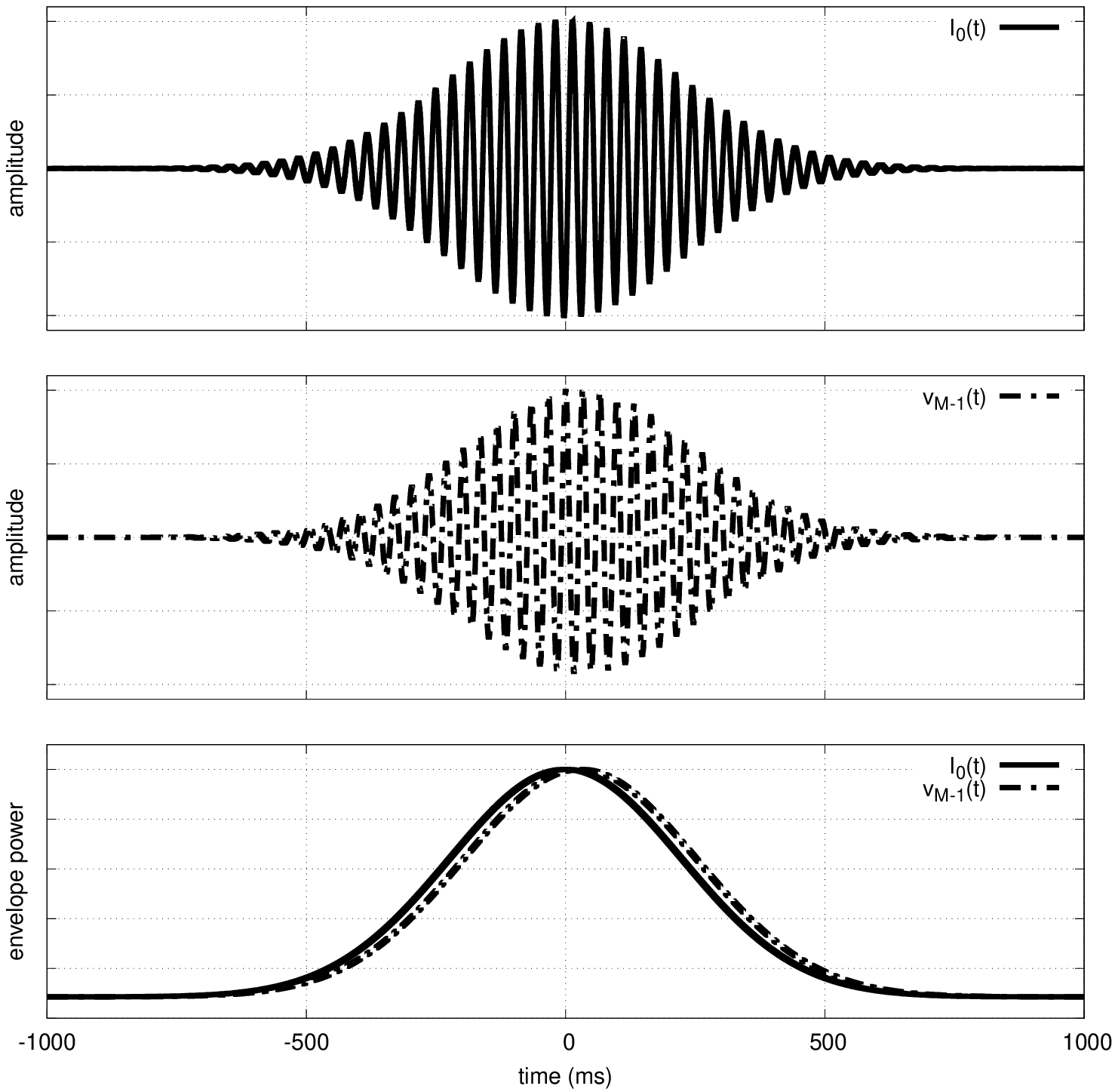}}
    \caption{\textbf{Anticipation and delay of Gaussian wave pulses} with carrier frequency within the negative and positive, respectively, group delay band by a chain of model neurons} \label{fig:wavepulse}
\end{figure}

The top panels of fig.\ref{fig:wavepulse} show the input wave pulses with carrier frequencies of $\nu_c=\nu_0/2\approx$\SI{7.57}{\hertz} (left-panel) and $\nu_c=2\nu_0\approx$\SI{30.28}{\hertz} (right-panel), falling, respectively, into the negative group delay band and a band near the maximal group delay of the membrane potential (c.f. Fig.\ref{fig:gd}).
The middle panels show the membrane potential of the last neuron in a chain of $17$ cascaded model neurons, in which the first neuron received the wave pulse as in the first panel
\begin{equation}
    I_0(t) = e^{-\alpha(t-t_0)^2}A\sin\left({\omega_c t}\right) \nonumber 
\end{equation}
at time $t=t_0$ with carrier frequency $\omega_c$ and amplitude $A=$\SI{1e-2}.

The bottom panels of Fig.\ref{fig:wavepulse} show the amplitude envelope of the input signal (solid lines) and that of the last neuron (dashed lines), measured as the output of a strong lowpass filter, for the two different signals. 

In the membrane potential of the last neuron, a clear shift forward in time of the signal envelope is visible for signals in the negative group delay band (bottom left panel), in agreement with the group delay predicted by (\ref{eq:gd}), whereas the envelope of the signal in the positive group delay band is delayed (bottom right panel). 

\begin{figure}
    \includegraphics[width=.9\columnwidth]{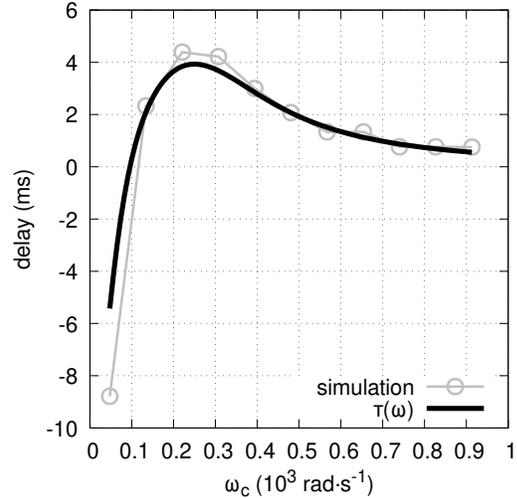}
    \caption{\textbf{Peak correlation time versus $\omega_c$} for Gaussian wave pulses. Numerical (connected dots) alongside theoretical (solid black line) group delay}\label{fig:wavepulse_shape}
\end{figure}
The time-shift versus pulse-carrier frequency is shown by fig.\ref{fig:wavepulse_shape} which shows the time lag (y-axis) of the peak in the correlation of the membrane potential of the first neuron, receiving an input Gaussian pulse with carrier frequency $\omega_c$ (x-axis), with the membrane potential of the last of the chain of $17$ neurons. It is visible that the numerical results (connected dots) agree well with the expected group delay $\tau(\omega_c)$ (solid black line). 

\subsection{Filtered noise}
Pure sinusoidal signals, even though they allow simple analyses, are rare. More commonly, biological and physical signals are considered to be noisy \citep[e.g.][]{Barkai2007, Ermentrout2008, Lindner2004, Selimkhanov2014, Tsimring2014}. In the following it is shown, as anticipated, that the membrane potential of the model neuron also anticipates the fluctuations of band-limited white noise input.

An ideal zero-mean Gaussian white noise $\xi(t)$ with variance $\sigma^2$ has a flat, wide-band, power spectral density $|\widetilde{\xi}(\omega)|^2 = \int \sigma^2 \delta(t)e^{-\mathit{i}\omega t}dt = \sigma^2$. 
In the following $\xi(t)$ is a normal Gaussian noise, thus $\sigma^2=1$. This signal will be filtered by a Gaussian filter centered at a frequency $\omega_c$ and a narrow band-width determined by $0<\alpha\ll 1$. Thus the first neuron receives an input with spectrum 
\begin{equation}
    \widetilde{I}_0(\omega) = A_0 e^{-\frac{(\omega-\omega_c)^2}{4\alpha}}  \widetilde{\xi}(\omega). \nonumber 
\end{equation}
Such narrow band filtering of sub-threshold inputs can arise due to signal transmission with heterogeneous transmission delays between neurons \citep{Houben2020a}.

\begin{figure}
    \subfloat[$\omega_c=\omega_0/2$\label{fig:bpnoise_ngd}]{\includegraphics[width=.5\columnwidth]{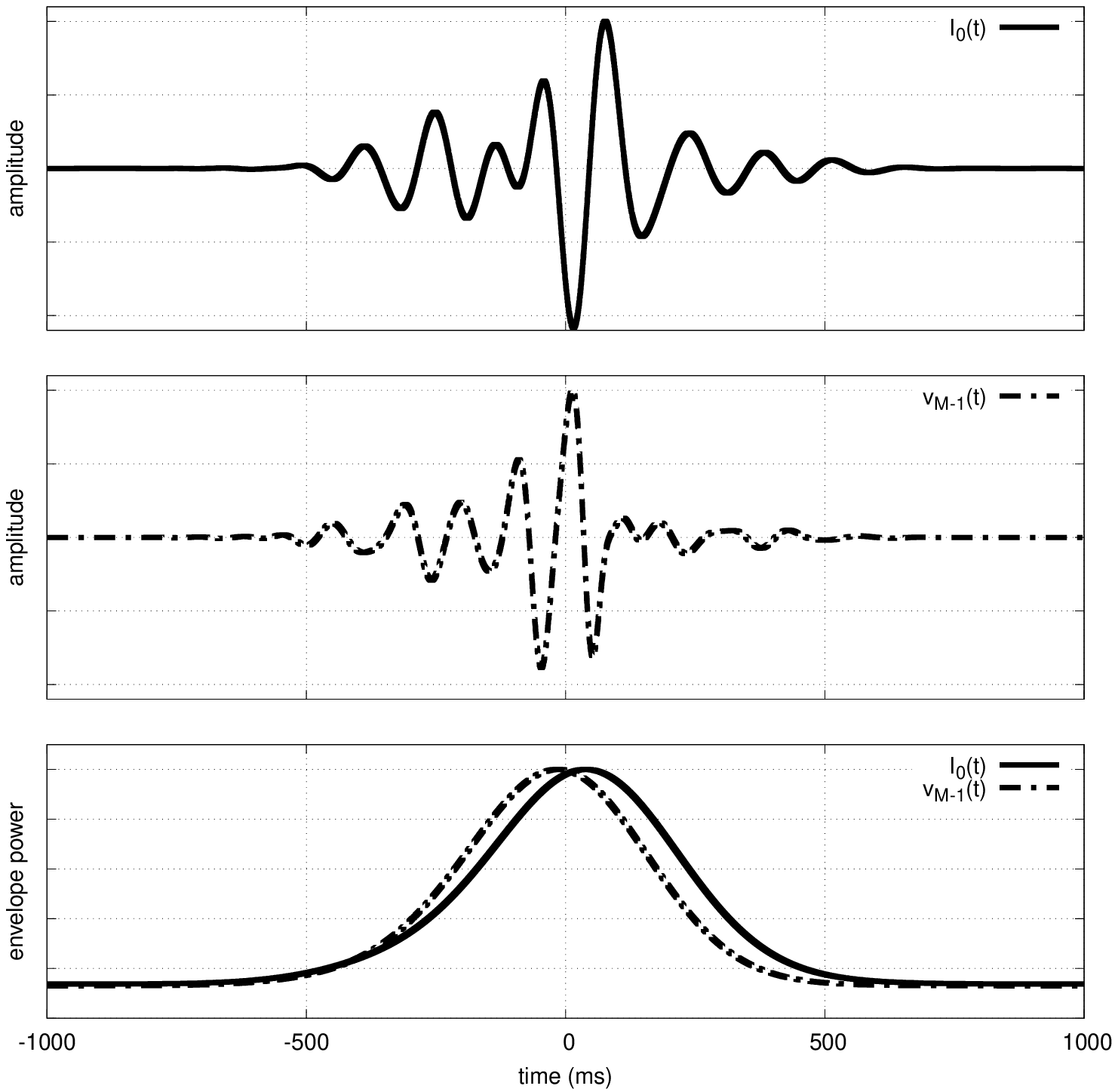}}
    \subfloat[$\omega_c=2\omega_0$\label{fig:bpnoise_pgd}]{\includegraphics[width=.5\columnwidth]{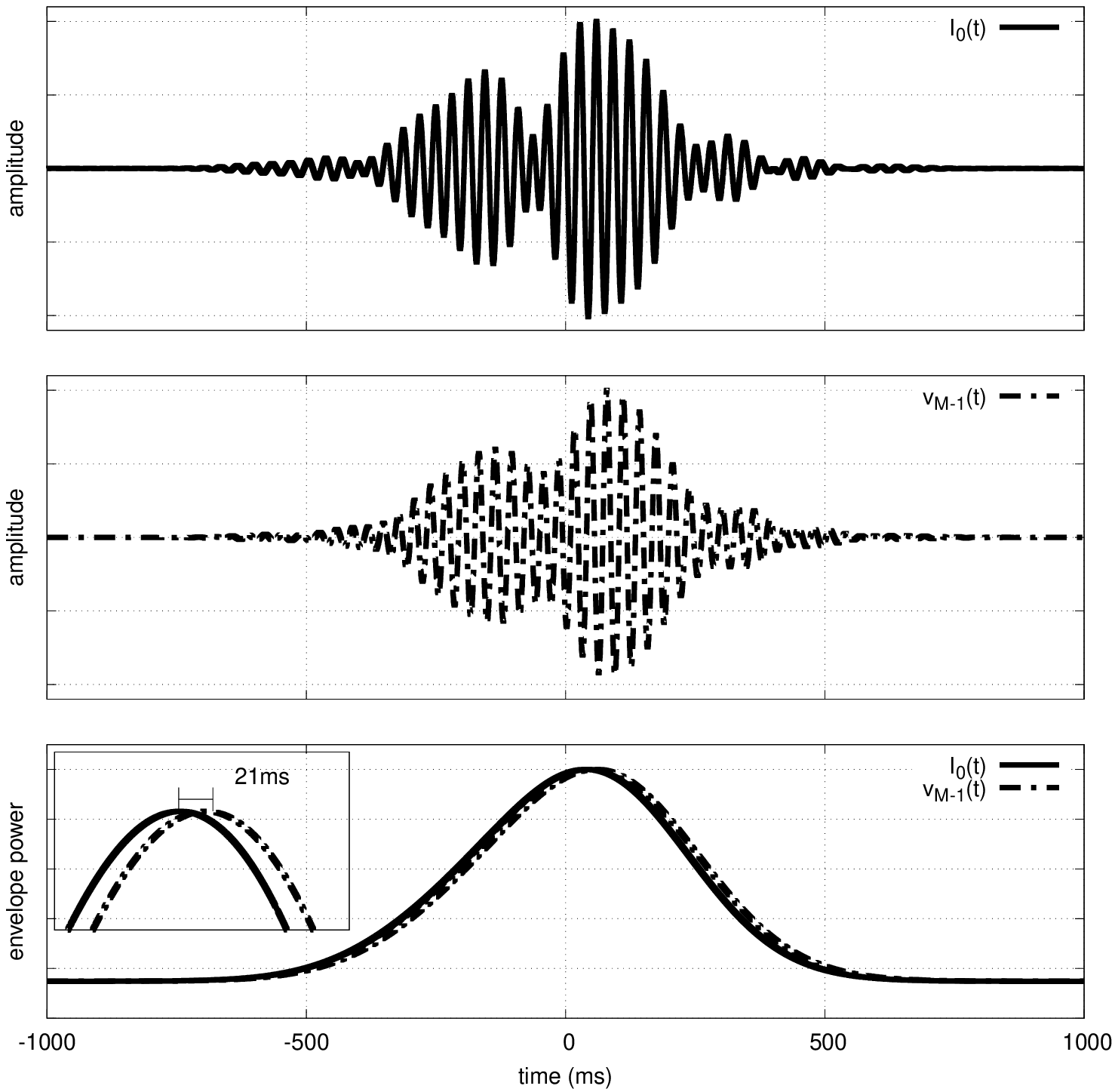}}
   %
    \caption{\textbf{Anticipation and delay of band-limited noise pulses} with pass-band within a negative and positive group delay band, respectively}\label{fig:bpnoise}
\end{figure}
The left column of fig.\ref{fig:bpnoise} shows the results of the chain of model neurons being driven by a band-limited noise with center-frequency $\nu_c=\nu_0/2\approx$ \SI{7.57}{\hertz}, and $\alpha=$\SI{1e-4} falling within the negative group delay band. 
The top panels show a section of one input signal (top most panel) together with the membrane potential of the last neuron in the chain (second to top panel)  in response to that signal. The bottom most panel shows the envelope of the input (solid line) and that of the membrane potential of the last neuron (dashed line). The right panel shows the same, but for an input with $\nu_c = 2\nu_0 \approx$ \SI{30.28}{\hertz}.
%

\begin{figure}
    \includegraphics[width=.9\columnwidth]{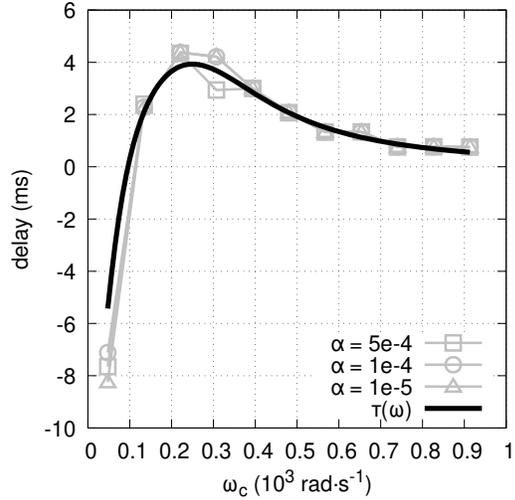}
    \caption{\textbf{Peak correlation time versus $\omega_c$} for filtered noise input. Numerical results (connected symbols) for different filtering band-widths, alongside theoretical group delay (solid black line)}\label{fig:bpnoise_shape}
\end{figure}
Fig.\ref{fig:bpnoise_shape} shows the time lag of the first peak in the correlation function between a filtered noise input with center-frequency $\omega_c$ (x-axis) and the membrane potential of the model neuron receiving the input (connected symbols), for different filtering bandwidths $\alpha$. The solid black line shows the theoretical group delay $\tau(\omega_c)$.

\subsection{Modulated \& filtered noise}
In the foregoing section the noise considered was stationary, however it is likely that in biological and physical settings the noise sources evolve over time. 
In addition, modulations of stationary noise sources can serve as a form of signalling through a medium. Such as mean- and variance-modulation as proposed `communication channels' for neurons and neural networks \citep[e.g.][]{Herfurth2019, Lindner2001}.

Considering a noise signal subjected to slow mean- and variance-modulation,  $g_\mu(t)$ and $g_\sigma(t)$, 
\begin{equation}
    f(t) = g_\mu(t) + g_\sigma(t)\xi(t), \nonumber
\end{equation}
with either the result $f(t)$ or the input noise $\xi(t)$ being passed through a band-pass filter as before. Clearly in each case the noise signal will be affected by the group delay, if the pass-band falls within a frequency range exhibiting a group delay.
It is also apparent that in both cases the mean-modulation $g_\mu(t)$ will not be affected by the group delay, and thus will be passed without delay or advance.

The variance-modulation however will be affected, but only in case of the noise signal being band-limited before the variance modulation, which is easily understood from the fact that the power of $g_\sigma(t)$ is focussed outside of the pass band of the band-limited filter and thus will only serve as an amplitude modulation to the higher frequency noise signal. 


\section{Conclusion}
In this paper it is shown that the dynamics of the voltage variable of the FitzHugh-Nagumo model posesses a negative group delay for slowly fluctuating inputs, by finding an expression for the group delay of a linear approximation of the governing equation for the voltage variable at the fixed point of the model system. 

The effects of the group delay are demonstrated numerically for different types of signals and signal modulations and it is shown that a chain of uni-directionally coupled model neurons anticipates certain aspects of inputs if the input consists of a carrier signal with a frequency falling within the negative group delay band.


\bibliographystyle{plain}

\begin{thebibliography}{10}

\bibitem{AstonJones1985}
G~Aston-Jones, SL~Foote, and M~Segal.
\newblock Impulse conduction properties of noradrenergic locus coeruleus axons
  projecting to monkey cerebrocortex.
\newblock {\em Neuroscience}, 15(3):765--777, 1985.

\bibitem{Bariska2008}
A~Bariska.
\newblock Time machine, anyone?, 2008.

\bibitem{Barkai2007}
Naama Barkai and Ben-Zion Shilo.
\newblock Variability and robustness in biomolecular systems.
\newblock {\em Molecular cell}, 28(5):755--760, 2007.

\bibitem{Bigelow2003}
Matthew~S Bigelow, Nick~N Lepeshkin, and Robert~W Boyd.
\newblock Superluminal and slow light propagation in a room-temperature solid.
\newblock {\em Science}, 301(5630):200--202, 2003.

\bibitem{Brillouin1914}
L{\'e}on Brillouin.
\newblock {\"U}ber die fortpflanzung des lichtes in dispergierenden medien.
\newblock {\em Annalen der Physik}, 349(10):203--240, 1914.

\bibitem{Buzsaki2006}
Gyorgy Buzsaki.
\newblock {\em Rhythms of the Brain}.
\newblock Oxford University Press, 2006.

\bibitem{Chiao1993}
Raymond~Y Chiao.
\newblock Superluminal (but causal) propagation of wave packets in transparent
  media with inverted atomic populations.
\newblock {\em Physical Review A}, 48(1):R34, 1993.

\bibitem{Ciszak2003}
Marzena Ciszak, O~Calvo, Cristina Masoller, Claudio~R Mirasso, and Ra{\'u}l
  Toral.
\newblock Anticipating the response of excitable systems driven by random
  forcing.
\newblock {\em Physical review letters}, 90(20):204102, 2003.

\bibitem{Cleland1976}
BG~Cleland, WR~Levick, R~Morstyn, and HG~Wagner.
\newblock Lateral geniculate relay of slowly conducting retinal afferents to
  cat visual cortex.
\newblock {\em The Journal of physiology}, 255(1):299--320, 1976.

\bibitem{Dayan2001}
Peter Dayan and Laurence~F Abbott.
\newblock {\em Theoretical neuroscience: computational and mathematical
  modeling of neural systems}.
\newblock Computational Neuroscience Series, 2001.

\bibitem{Delorme2003}
Arnaud Delorme.
\newblock Early cortical orientation selectivity: how fast inhibition decodes
  the order of spike latencies.
\newblock {\em Journal of computational neuroscience}, 15(3):357--365, 2003.

\bibitem{Ermentrout2008}
G~Bard Ermentrout, Roberto~F Gal{\'a}n, and Nathaniel~N Urban.
\newblock Reliability, synchrony and noise.
\newblock {\em Trends in neurosciences}, 31(8):428--434, 2008.

\bibitem{FitzHugh1961}
Richard FitzHugh.
\newblock Impulses and physiological states in theoretical models of nerve
  membrane.
\newblock {\em Biophysical journal}, 1(6):445, 1961.

\bibitem{FourcaudTromce2003}
Nicolas Fourcaud-Trocm{\'e}, David Hansel, Carl Van~Vreeswijk, and Nicolas
  Brunel.
\newblock How spike generation mechanisms determine the neuronal response to
  fluctuating inputs.
\newblock {\em Journal of Neuroscience}, 23(37):11628--11640, 2003.

\bibitem{Garrett1970}
CGB Garrett and DE~McCumber.
\newblock Propagation of a gaussian light pulse through an anomalous dispersion
  medium.
\newblock {\em Physical Review A}, 1(2):305, 1970.

\bibitem{Gray1989}
Charles~M Gray, Peter K{\"o}nig, Andreas~K Engel, and Wolf Singer.
\newblock Oscillatory responses in cat visual cortex exhibit inter-columnar
  synchronization which reflects global stimulus properties.
\newblock {\em Nature}, 338(6213):334--337, 1989.

\bibitem{Herfurth2019}
Tim Herfurth and Tatjana Tchumatchenko.
\newblock Information transmission of mean and variance coding in
  integrate-and-fire neurons.
\newblock {\em Physical Review E}, 99(3):032420, 2019.

\bibitem{Houben2020a}
Akke~Mats Houben.
\newblock Frequency selectivity of neural circuits with heterogeneous
  transmission delays.
\newblock {\em arXiv preprint arXiv:2009.09250}, 2020.

\bibitem{Izhikevich2007}
Eugene~M Izhikevich.
\newblock {\em Dynamical systems in neuroscience}.
\newblock MIT press, 2007.

\bibitem{KoksalErsoz2019}
Elif K{\"o}ksal~Ers{\"o}z, Mathieu Desroches, Claudio~R Mirasso, and Serafim
  Rodrigues.
\newblock Anticipation via canards in excitable systems.
\newblock {\em Chaos: An Interdisciplinary Journal of Nonlinear Science},
  29(1):013111, 2019.

\bibitem{Kuramoto1975}
Yoshiki Kuramoto.
\newblock Self-entrainment of a population of coupled non-linear oscillators.
\newblock In {\em International symposium on mathematical problems in
  theoretical physics}, pages 420--422. Springer, 1975.

\bibitem{Lindner2004}
Benjamin Lindner, Jordi Garc{\i}a-Ojalvo, Alexander Neiman, and Lutz
  Schimansky-Geier.
\newblock Effects of noise in excitable systems.
\newblock {\em Physics reports}, 392(6):321--424, 2004.

\bibitem{Lindner2001}
Benjamin Lindner and Lutz Schimansky-Geier.
\newblock Transmission of noise coded versus additive signals through a
  neuronal ensemble.
\newblock {\em Physical Review Letters}, 86(14):2934, 2001.

\bibitem{Lucyszyn1993}
S~Lucyszyn, ID~Robertson, and AH~Aghvami.
\newblock Negative group delay synthesiser.
\newblock {\em Electronics Letters}, 29(9):798--800, 1993.

\bibitem{Matias2014}
Fernanda~S Matias, Leonardo~L Gollo, Pedro~V Carelli, Steven~L Bressler, Mauro
  Copelli, and Claudio~R Mirasso.
\newblock Modeling positive granger causality and negative phase lag between
  cortical areas.
\newblock {\em NeuroImage}, 99:411--418, 2014.

\bibitem{Mitchell1997}
Morgan~W Mitchell and Raymond~Y Chiao.
\newblock Negative group delay and ``fronts'' in a causal system: An experiment
  with very low frequency bandpass amplifiers.
\newblock {\em Physics Letters A}, 230(3-4):133--138, 1997.

\bibitem{Nagumo1962}
Jinichi Nagumo, Suguru Arimoto, and Shuji Yoshizawa.
\newblock An active pulse transmission line simulating nerve axon.
\newblock {\em Proceedings of the IRE}, 50(10):2061--2070, 1962.

\bibitem{Nakanishi2002}
T~Nakanishi, K~Sugiyama, and M~Kitano.
\newblock Demonstration of negative group delays in a simple electronic
  circuit.
\newblock {\em American Journal of Physics}, 70(11):1117--1121, 2002.

\bibitem{Naundorf2005}
B~Naundorf, Theo Geisel, and Fred Wolf.
\newblock Action potential onset dynamics and the response speed of neuronal
  populations.
\newblock {\em Journal of computational neuroscience}, 18(3):297--309, 2005.

\bibitem{Oram1992}
MW~Oram and DI~Perrett.
\newblock Time course of neural responses discriminating different views of the
  face and head.
\newblock {\em Journal of neurophysiology}, 68(1):70--84, 1992.

\bibitem{Panzeri1996}
S~Panzeri, G~Biella, ET~Rolls, WE~Skaggs, and A~Treves.
\newblock Speed, noise, information and the graded nature of neuronal
  responses.
\newblock {\em Network: Computation in Neural Systems}, 7(2):365--370, 1996.

\bibitem{Pyragiene2013}
T~Pyragien{\.e} and K~Pyragas.
\newblock Anticipating spike synchronization in nonidentical chaotic neurons.
\newblock {\em Nonlinear Dynamics}, 74(1-2):297--306, 2013.

\bibitem{vanRullen2005}
Rufin~van Rullen, Rudy Guyonneau, and Simon~J Thorpe.
\newblock Spike times make sense.
\newblock {\em Trends in neurosciences}, 28(1):1--4, 2005.

\bibitem{Segard1985}
Bernard Segard and Bruno Macke.
\newblock Observation of negative velocity pulse propagation.
\newblock {\em Physics Letters A}, 109(5):213--216, 1985.

\bibitem{Selimkhanov2014}
Jangir Selimkhanov, Brooks Taylor, Jason Yao, Anna Pilko, John Albeck,
  Alexander Hoffmann, Lev Tsimring, and Roy Wollman.
\newblock Accurate information transmission through dynamic biochemical
  signaling networks.
\newblock {\em Science}, 346(6215):1370--1373, 2014.

\bibitem{Sommerfeld1914}
Arnold Sommerfeld.
\newblock {\"U}ber die fortpflanzung des lichtes in dispergierenden medien.
\newblock {\em Annalen der Physik}, 349(10):177--202, 1914.

\bibitem{Stenner2003}
Michael~D Stenner, Daniel~J Gauthier, and Mark~A Neifeld.
\newblock The speed of information in a ‘fast-light’optical medium.
\newblock {\em Nature}, 425(6959):695--698, 2003.

\bibitem{Strogatz2000}
Steven~H Strogatz.
\newblock From kuramoto to crawford: exploring the onset of synchronization in
  populations of coupled oscillators.
\newblock {\em Physica D: Nonlinear Phenomena}, 143(1-4):1--20, 2000.

\bibitem{Swadlow1974}
Harvey~A Swadlow.
\newblock Systematic variations in the conduction velocity of slowly conducting
  axons in the rabbit corpus callosum.
\newblock {\em Experimental neurology}, 43(2):445--451, 1974.

\bibitem{Swadlow1985}
Harvey~A Swadlow and Theodore~G Weyand.
\newblock Receptive-field and axonal properties of neurons in the dorsal
  lateral geniculate nucleus of awake unparalyzed rabbits.
\newblock {\em Journal of neurophysiology}, 54(1):168--183, 1985.

\bibitem{Tchumatchenko2011}
Tatjana Tchumatchenko, Aleksey Malyshev, Fred Wolf, and Maxim Volgushev.
\newblock Ultrafast population encoding by cortical neurons.
\newblock {\em Journal of Neuroscience}, 31(34):12171--12179, 2011.

\bibitem{Thorpe1996}
Simon Thorpe, Denis Fize, and Catherine Marlot.
\newblock Speed of processing in the human visual system.
\newblock {\em nature}, 381(6582):520--522, 1996.

\bibitem{Tovee1994}
Martin~J Tov{\'e}e.
\newblock Neuronal processing: How fast is the speed of thought?
\newblock {\em Current Biology}, 4(12):1125--1127, 1994.

\bibitem{Tsimring2014}
Lev~S Tsimring.
\newblock Noise in biology.
\newblock {\em Reports on Progress in Physics}, 77(2):026601, 2014.

\bibitem{Varela2001}
Francisco Varela, Jean-Philippe Lachaux, Eugenio Rodriguez, and Jacques
  Martinerie.
\newblock The brainweb: phase synchronization and large-scale integration.
\newblock {\em Nature reviews neuroscience}, 2(4):229--239, 2001.

\bibitem{Voss2000}
Henning~U Voss.
\newblock Anticipating chaotic synchronization.
\newblock {\em Physical review E}, 61(5):5115, 2000.

\bibitem{Voss2016}
Henning~U Voss.
\newblock The leaky integrator with recurrent inhibition as a predictor.
\newblock {\em Neural Computation}, 28(8):1498--1502, 2016.

\bibitem{Wang2000}
Lijun~J Wang, A~Kuzmich, and Arthur Dogariu.
\newblock Gain-assisted superluminal light propagation.
\newblock {\em Nature}, 406(6793):277--279, 2000.

\bibitem{Winfree1967}
Arthur~T Winfree.
\newblock Biological rhythms and the behavior of populations of coupled
  oscillators.
\newblock {\em Journal of theoretical biology}, 16(1):15--42, 1967.

\bibitem{Woodley2004}
JF~Woodley and M~Mojahedi.
\newblock Negative group velocity and group delay in left-handed media.
\newblock {\em Physical Review E}, 70(4):046603, 2004.

\end{thebibliography}

\end{document}